\documentclass[floats,floatfix,amssymb,prd,twocolumn,superscriptaddress,nofootinbib]{revtex4-2}

\usepackage{amssymb,amsmath,verbatim,mathtools,needspace,enumitem,etoolbox,graphicx,physics,microtype,afterpage,bm}
\usepackage[dvipsnames, usenames]{xcolor}
\definecolor{linkcolor}{rgb}{0.0,0.3,0.5}
\usepackage[unicode, colorlinks=true, linkcolor=linkcolor, citecolor=linkcolor, filecolor=linkcolor,urlcolor=linkcolor, pdfusetitle]{hyperref}
\usepackage[all]{hypcap}
\usepackage[T1]{fontenc}
\usepackage[utf8]{inputenc}
\usepackage{tabularx}
\usepackage{float}
\usepackage{lmodern}

\allowdisplaybreaks

\usepackage{tikz}
\usepackage{color}
\usepackage{framed}
\usepackage{hyperref}
\hypersetup{colorlinks, citecolor=bluscuro, linkcolor=black, urlcolor=bluscuro}

\usepackage{bm}
\usepackage{latexsym}
\usepackage{amsmath}
\usepackage{amssymb}
\usepackage{amsthm}
\usepackage{amscd}
\usepackage[mathscr]{eucal}
\usepackage{graphicx}
\usepackage{subfigure}
\usepackage{psfrag}
\usepackage{rotating}

\usepackage{color}

\definecolor{ggreen}{cmyk}{0.7,     0,      0.9,      0}
\definecolor{viol}{cmyk}{0.3,1,0,0}
\definecolor{myred}{cmyk}{0.1, 1, 0.5, 0}
\definecolor{bblue}{rgb}{0.2, 0.29996, 0.8 }

\definecolor{rossos}{cmyk}{0,1,1,0.55}
\definecolor{bluscuro}{rgb}{0.15, 0.2, .85}
\definecolor{bluchiaro}{cmyk}{1,.3,0.,0.1}

\definecolor{ForestGreen}{rgb}{0.13, 0.55, 0.13}

\theoremstyle{plain}

\begin{document}


\title{Compact elastic objects in general relativity}

\author{Artur Alho}
\affiliation{Center for Mathematical Analysis, Geometry and Dynamical Systems, Instituto Superior T\'ecnico, Universidade de Lisboa, Av. Rovisco Pais, 1049-001 Lisboa, Portugal}

\author{Jos\'e Nat\'ario}
\affiliation{Center for Mathematical Analysis, Geometry and Dynamical Systems, Instituto Superior T\'ecnico, Universidade de Lisboa, Av. Rovisco Pais, 1049-001 Lisboa, Portugal}

\author{Paolo Pani}
\affiliation{Dipartimento di Fisica, Sapienza Universit\`a di Roma \& INFN Roma1, Piazzale Aldo Moro 5, 00185, Roma, Italy}

\author{Guilherme Raposo}
\affiliation{CENTRA, Instituto Superior T\'ecnico, Universidade de Lisboa, Av. Rovisco Pais, 1049-001 Lisboa, Portugal}
\affiliation{Centre for Research and Development in Mathematics and Applications (CIDMA), Campus de Santiago, 3810-183 Aveiro, Portugal}
\affiliation{Dipartimento di Fisica, Sapienza Universit\`a di Roma \& INFN Roma1, Piazzale Aldo Moro 5, 00185, Roma, Italy}


\begin{abstract}
We introduce a rigorous and general framework to study systematically self-gravitating elastic materials within general relativity, and apply it to investigate the existence and viability, including radial stability, of spherically symmetric elastic stars.
We present the mass-radius ($M-R$) diagram for various families of models, showing that elasticity contributes to increase the maximum mass and the compactness up to $\approx 22\%$, thus supporting compact stars with mass well above two solar masses. Some of these elastic stars can reach compactness as high as $GM/(c^2R)\approx 0.35$ while remaining stable under radial perturbations and satisfying all energy conditions and subluminal wave propagation, thus being physically realizable models of stars with a light ring.
We provide numerical evidence that radial instability occurs for central densities larger than that corresponding to the maximum mass, as in the perfect fluid case.
Elasticity may be a key ingredient to build consistent models of exotic ultracompact objects and black-hole mimickers, and can also be relevant for a more accurate modelling of the interior of neutron stars.
\end{abstract}

\maketitle

\noindent{{\bf{\em Introduction.}}}
%
Astronomical compact objects are typically idealized as self-gravitating (often perfect) fluids, wherein (isotropic) pressure prevents gravitational collapse.
However, while degenerate fermions behave as a weakly-interacting gas at relatively small densities, nuclear interactions and QCD effects become crucial inside relativistic stars. Thus, it is reasonable to expect that the perfect fluid idealization will eventually break down, at least to some extent, and that solid phases of matter may be relevant for astronomical compact objects. This is indeed the situation in the crust of a neutron star~\cite{Chamel:2008ca,Suleiman:2021hre}, whose fundamental constituents are largely unknown, especially in the core~\cite{Lattimer:2004pg}.

A natural generalization of fluid models is to consider elastic materials~\cite{CarterQuintana,Par00,Karlovini:2002fc,FK07,Andreasson:2014lka}, 
%
also studied perturbatively to model the crust of a neutron star~\cite{Chamel:2008ca,Suleiman:2021hre}.
In this letter, we introduce a  new systematic approach to the problem of self-gravitating elastic materials in General-Relativity~(GR), which 
allows building elastic compact objects in a simple --~yet general~-- way and assessing their viability in the strong gravity regime.

Beside offering a more accurate description of the stellar interior~\cite{Rajagopal_2006,Rajagopal_2006b,Alford_2008}, elasticity might play a crucial role in constructing consistent models of exotic compact objects and black-hole mimickers within GR and extensions thereof~\cite{Cardoso:2019rvt,Carballo-Rubio:2018jzw}. Under certain hypotheses~\cite{Cardoso:2019rvt,Urbano:2018nrs}, Buchdhal's theorem~\cite{Buchdahl:1959zz} states that self-gravitating, perfect fluid GR solutions satisfy the following bound on the compactness: $M/R\leq4/9$, where $M$ is the mass and $R$ is the radius of the star (henceforth we use $G=c=1$ units). 
Buchdhal's theorem assumes that matter is described by a perfect fluid, and can be extended to mildly anisotropic fluids for which the radial pressure is larger than the tangential one~\cite{Urbano:2018nrs} (see~\cite{Andreasson:2007ck,Karageorgis:2007cy} for more general results). Indeed, compact objects made of strongly anisotropic fluids (e.g., gravastars~\cite{Mazur:2004fk} and anisotropic stars~\cite{Raposo:2018rjn}) can have higher compactness and a continuous BH limit, $M/R\to1/2$~\cite{Pani:2015tga,Uchikata:2015yma,Uchikata:2016qku,Beltracchi:2021lez}. However, the viability of such ultracompact models is questionable, since they either violate some of the energy conditions~\cite{HawkingEllis}, or feature superluminal speed of sound or ad-hoc thin-shells within the fluid (see \cite{Cardoso:2019rvt} for a discussion). On the other hand, physically realizable models like boson stars 
are not significantly more compact than an ordinary perfect fluid neutron star in the static case~\cite{Liebling:2012fv}.

Since elastic materials feature shears and anisotropies, it is natural to ask whether full-fledged, physically realizable models of \emph{ultracompact}~\cite{Cardoso:2019rvt} elastic stars can be built. In this letter we will show that this is the case. Viable elastic stars can have $M/R>1/3$, thus featuring the same Schwarzschild photon sphere at radial coordinate $r=3M$, a crucial property to mimic the phenomenology of black holes~\cite{Cardoso:2014sna,Cunha:2017eoe,Carballo-Rubio:2018jzw,Cardoso:2019rvt}. We will also show that elastic stars can exceed the Buchdhal's bound on the compactness, but only in their unstable or superluminal branch, at least for the class of materials under consideration.

\noindent{{\bf{\em Setup.}}}
%
We focus on spherical symmetry and study both static solutions and their dynamical radial perturbations. More details and models will be given in a companion paper~\cite{inprep}. In Schwarzschild coordinates, the line element reads   $ds^2=-e^{2\alpha(t,r)}dt^2+e^{2\beta(t,r)}dr^2+r^2 d\Omega^2$, where $d\Omega^2$ is the metric of the unit $2$-sphere. A spherically symmetric self-gravitating body is described in terms of the scalars $(\rho,p_\mathrm{rad},p_\mathrm{tan},v)$ satisfying the Einstein's equations, where $\rho(t,r)$ is the energy density, $p_\mathrm{rad}(t,r)$ and $p_\mathrm{tan}(t,r)$ are the radial and tangential pressures, respectively, and $v(t,r)$ is the radial velocity. The 4-velocity of matter is $u^{\mu}=(e^{-\alpha}\langle v\rangle,v,0,0)$, where $\langle v\rangle = \sqrt{1+e^{2\beta}v^2}$. The Einstein equations are closed by postulating equations of state~(EoS) relating the pressures and the density.  

Relativistic elasticity is based on a variational principle wherein the Lagrangian density is covariant under spacetime diffeomorphisms, and consists of the sum of the rest-frame energy density for the undeformed material and a (deformation) potential energy density, so that it coincides with the total energy density $\rho$ measured by an observer at rest with respect to the material~\cite{Beig:2002pk,Brown:2020pav}. For homogeneous and isotropic elastic materials and under spherically symmetry, the Lagrangian is given by~\cite{inprep}
\begin{equation}\label{ED}
\widehat{\rho}({{\delta}},\eta) =\delta ( \rho_0  +\widehat{w}({{\delta}},\eta) )\,,
\end{equation}
where the potential energy density (which we will call stored energy function, by analogy with the Newtonian case) is $w(t,r)=\widehat{w}(\delta(t,r),\eta(t,r))$, so that $\rho(t,r)=\widehat{\rho}(\delta(t,r),\eta(t,r))$. Here
\begin{align}
\delta(t,r)&=\frac{n(t,r)}{n_0},\\
\eta(t,r)&=\frac{3}{r^3}\int^r_0 e^{\beta(t,u)}\langle v(t,u)\rangle{{\delta}}(t,u)u^2 du\,, 
\end{align}
where $n(t,r)$ is the (conserved) particle number density in the physical (deformed) state, and $n_0>0$ and $\rho_0>0$ are the particle number density and energy density in the reference material frame, respectively. The reference state is an idealized state with $n=n_0$, $\beta=0$ (corresponding to a flat material metric), and $v=0$, that is, $({{\delta}},\eta)=(1,1)$. 
The EoS are 
\begin{subequations}
	\begin{align}
	&\widehat{p}_{\mathrm{rad}}({{\delta}},\eta)={{\delta}}\partial_{{\delta}}\widehat{\rho}({{\delta}},\eta)\, -\widehat{\rho}({{\delta}},\eta)\,,  
	\\
	&\widehat{q}(\delta,\eta)\equiv\widehat{p}_{\mathrm{tan}}({{\delta}},\eta)-\widehat{p}_{\mathrm{rad}}({{\delta}},\eta)=\frac{3}{2}\eta \partial_{\eta}\widehat{\rho}({{\delta}},\eta) \,.
	\end{align}	\label{EoS}
\end{subequations}	
Note that the perfect fluid case is included for any Lagrangian such that $\partial_\eta\widehat\rho=0$ (equivalently $\partial_\eta \widehat{w}=0$).

The function $\widehat{w}({{\delta}},\eta)$ satisfies the natural reference state condition, $\widehat{w}(1,1)=0$ (state of zero energy), i.e.,
\begin{equation}\label{NSC}
\widehat{\rho}(1,1)=\rho_0.
\end{equation}
The radial and tangential pressures satisfy the reference state condition
\begin{equation}\label{RefP}
\widehat{p}_{\mathrm{rad}}(1,1)=	\widehat{p}_{\mathrm{tan}}(1,1)=p_0\,.
\end{equation}
The reference state is said to be stress-free (pre-stressed) if $p_0=0$ ($p_0\neq0$). Furthermore, compatibility with linear elasticity requires
\begin{subequations}\label{LE}
	\begin{align}
	&\partial_\delta\widehat{p}_{\mathrm{rad}}(1,1)=\lambda+2\mu ,\qquad \partial_\eta \widehat{p}_{\mathrm{rad}}(1,1) =-\frac{4}{3}\mu\,, \\
	&\partial_\delta\widehat{p}_{\mathrm{tan}}(1,1)=\lambda ,\qquad\qquad\, \partial_\eta \widehat{p}_{\mathrm{tan}}(1,1) =\frac{2}{3}\mu\,,
	\end{align}
\end{subequations}
where $\lambda$, $\mu$ are the Lam\'e parameters. 

Restricting to static configurations, the radial velocity vanishes ($v=0$), while $(\rho,p_\mathrm{rad},p_\mathrm{tan})$ and $\alpha$, $\beta$ are functions of the areal coordinate $r$ only. In this case the Einstein equations reduce to the Tolman-Oppenheimer-Volkoff (TOV) equations 
\begin{align}
\frac{dp_\mathrm{rad}}{dr} &=\frac{2}{r}(p_\mathrm{tan}-p_\mathrm{rad})-(p_\mathrm{rad}+\rho)\frac{d\alpha}{dr}\,,\label{TOVeq} \\
\frac{d\alpha}{dr} &= \frac{e^{2\beta}}{r}\left(\frac{m}{r}+{4\pi  r^2}p_{\mathrm{rad}}\right)\,,
\end{align}	
%
where $e^{-2\beta(r)}=1-\frac{2m(r)}{r}$, $\alpha(r)$ is the relativistic gravitational potential, and $m(r)=4\pi\int^{r}_0\rho(u)u^2 du$ is the Misner-Sharp mass.
%

In terms of the variables $(\delta(r),\eta(r),m(r))$, the TOV equations become a closed first-order system,
\begin{subequations}\label{TOV2}
	\begin{align}
	&\partial_{{{\delta}}}\widehat{p}_{\mathrm{rad}} \frac{d{{\delta}}}{dr} = \frac{3}{r}\partial_{\eta}\widehat{p}_{\mathrm{rad}}(\eta-e^{\beta}{{\delta}})
	+\frac{2}{r}\widehat{q}\nonumber\\
	&\qquad-\left(\widehat{p}_{\mathrm{rad}}+\widehat{\rho}\right) \frac{e^{2\beta}}{r}\left(\frac{m}{r} +4\pi  r^2 \widehat{p}_{\mathrm{rad}}\right)\,, \\
	&\frac{d\eta}{dr} =-\frac{3}{r}(\eta-e^{\beta}{{\delta}}) \,,\qquad \frac{dm}{dr}=4\pi r^2 \widehat{\rho} \,,
	\end{align} 
\end{subequations}
supplemented by~\eqref{ED} and the EoS~\eqref{EoS}. In the Newtonian limit we recover the results of~\cite{Alho:2018mro,Alho:2019fup}. For regular solutions we have $\lim\limits_{r\rightarrow0^+}{{\delta}}(r)={{\delta}}_c$ and $\lim\limits_{r\rightarrow 0^+} e^{\beta(r)}=1$, implying $\lim\limits_{r\rightarrow 0^+} \eta(r)={{\delta}}_c$. Regularity at the origin imposes $p_\mathrm{rad}(0)=p_\mathrm{tan}(0)$, which in turn implies
\begin{equation}\label{ISO}
\widehat{q}(\delta,\delta)=0 \qquad \text{for all}\quad {{\delta}}>0.
\end{equation}
%

\noindent{{\bf{\em Material models.}}}
The above formalism is general for any given set of functions $(\rho,p_\mathrm{rad},p_\mathrm{tan})$ satisfying Eqs.~\eqref{NSC},~\eqref{RefP},~\eqref{LE}, and~\eqref{ISO}. Here we
focus on elastic constitutive functions that are continuous deformations of relativistic polytropes (other examples are given in~\cite{inprep}). The simplest model is 
\begin{equation}\label{QM}
\widehat{\rho}(\delta,\eta) = (1-\kappa n) \rho_0 \delta + \kappa n \rho_0 \delta^{1+\frac1n} + \varepsilon \rho_0 (\delta-\eta)^2,
\end{equation}
which contains three dimensionless parameters $\kappa$, $n$ and $\varepsilon$. For $\varepsilon=0$ we have a perfect fluid with a polytropic EoS with polytropic index $n$ and reference state pressure $p_0=\kappa\rho_0>0$. 
The first Lam\'{e} parameter is $\lambda = \kappa \rho_0 \left( 1 + \frac1n \right) - \varepsilon \rho_0$, and the shear modulus is $\mu = \frac32 \varepsilon \rho_0$, which implies $\varepsilon\geq0$.

\begin{figure*}[t]
	\centering
	\label{fig:MR}
	\includegraphics[width=0.85\textwidth]{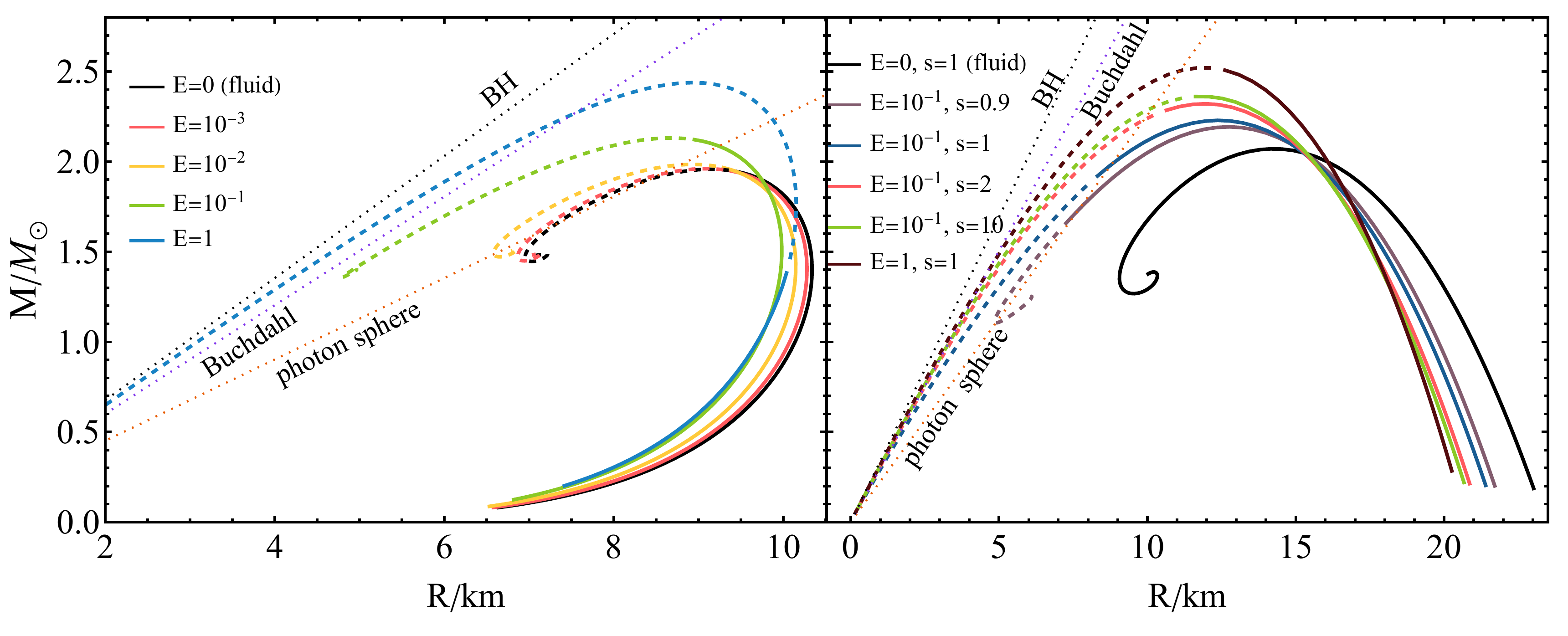} \\
	\caption{Mass-radius diagram for the quadratic elastic model~\eqref{QM} with $n=1/2$ and $K=6\times 10^4 M_\odot^4$
(left) and for the two-parameter elastic model~\eqref{PolyStore2} with $n=1$ and $K=160 M_\odot^2$ (right). Solid (dashed) curves correspond to configurations with subluminal (superluminal) wave propagation. For some regions of the parameter space (cf. green curve of left panel) there are configurations below the maximum mass featuring a photon sphere ($R<3M$) and subluminal wave propagation.
	}
\end{figure*}

The Newtonian limit of the previous quadratic model leads to equations of motion which are invariant under homologous transformations~\cite{Cha39} only for $n=1$. It is possible to generalize this stored energy function to one that leads to Newtonian equations of motion which are invariant under homologous transformations for general polytropic index $n$~\cite{inprep}:
\begin{equation}\label{PolyStore2}
\begin{split}
\widehat{w}(\delta,\eta)=&w_0 
+\eta^{\frac{1}{n}}\left[w_1+w_2\left(\frac{\delta}{\eta}\right)^{-1}+w_3\left(\frac{\delta}{\eta}\right)^{\frac{1}{s}}\right]\,, \\
\end{split}
\end{equation}
where $w_0 \equiv-n\kappa\rho_0$, $w_1\equiv \frac{(n-s)(1+n)}{n}\rho_0\kappa-2s\rho_0\varepsilon$, $w_2\equiv \frac{(s-n)}{(1+s)n}\rho_0\kappa+\frac{2s}{1+s}\rho_0\varepsilon$, $w_3\equiv \frac{(1+n)s^2}{(1+s)n}\rho_0\kappa+\frac{2s^2}{1+s}\rho_0\varepsilon$.
Here $s$ can be interpreted as the shear index; when $s=n$ and $\varepsilon=0$ we recover the usual relativistic polytropes (see also~\cite{ACL21,Cal21} for similar stored energy functions in the Newtonian setting). 
The Lam\'e parameters are the same as in the quadratic model~\eqref{QM}, and in fact the two models coincide when  $s=n=1$. 

When the model is stress-free, $p_0=0$, the reference state $(\delta,\eta)=(1,1)$ is uniquely defined. 
However, pre-stressed models (such as the ones we are considering) do not have a preferred reference state. In this case a different reference state, compressed or expanded with respect to the original reference state, provides an equivalent description of the material, moving from the parameters $(\rho_0,\kappa,\varepsilon)$ to new parameters $(\tilde\rho_0,\tilde\kappa,\tilde\varepsilon)$. The choice of reference state is thus akin to a gauge choice~\cite{inprep}.

It can be shown that in the fluid case $\widehat{p}_{\text{rad}} = \widehat{p}_{\text{tan}} = \kappa \rho_0 \delta^{1+\frac1n} = K {\widehat\sigma}^{1+\frac1n}$,
where $\widehat\sigma$ is the baryon density and $K = {\kappa}(1-\kappa n)^{-\frac{n+1}{n}} \rho_0^{-\frac1n}$.
The latter quantity is in fact invariant under renormalization of the reference state.
Moreover, changing $K$ only changes the mass scale of equilibrium configuration, and does not affect the value of dimensionless ratios such as the compactness.
Another invariant quantity under renormalization is
\begin{equation}
E = \frac{\varepsilon}{\kappa}\left(\frac{\kappa}{1-\kappa n}\right)^{1-n} \quad \text{or} \qquad E = \frac{\varepsilon}{\kappa} \,,
\end{equation}
in the case of model~\eqref{QM} or model~\eqref{PolyStore2}, respectively. 

\noindent{{\bf{\em Equilibrium configurations.}}}
%
The equations~\eqref{TOV2} for the stellar structure should be solved by requiring regularity of the functions at the center of the star. The solutions form a one-parameter family in terms of $\delta_c$, or, equivalently, of the central density $\rho(0)=\widehat\rho(\delta_c,\delta_c)=\rho_c$.
The radius $R$ of the star is defined by the condition $p_\mathrm{rad}(R)=0$, whereas $\rho(R)$ and $p_\mathrm{tan}(R)$ do not need to vanish.
Due to Birkhoff's theorem, the metric at $r>R$ (where $\rho=p_\mathrm{rad}=p_\mathrm{tan}=0$) is the standard Schwarzschild metric with $m(R)=M$ and $\alpha(r)=-\beta(r)$.
%

Within GR, physically viable matter fields should satisfy the following energy conditions~\cite{HawkingEllis}: 
\begin{subequations}
	\begin{align}
	\mathrm{SEC}&:\, \rho+p_\mathrm{rad}+2p_\mathrm{tan}\geq 0;\quad \rho+p_\mathrm{rad}\geq 0; \quad \rho+p_\mathrm{tan}\geq 0; \nonumber\\
	\mathrm{WEC}&:\, \rho\geq 0;\quad \rho+p_\mathrm{rad}\geq 0; \quad \rho+p_\mathrm{tan}\geq 0; \nonumber\\
	\mathrm{NEC}&:\, \rho+p_\mathrm{rad}\geq 0; \quad \rho+p_\mathrm{tan}\geq 0; \nonumber\\
	\mathrm{DEC}&:  \rho\geq |p_\mathrm{rad}|; \quad \rho\geq |p_\mathrm{tan}|, \nonumber
	\end{align}
\end{subequations}
for the strong, weak, null, and dominant energy condition, respectively.
Some further restrictions come from requiring causal wave propagation within the material. 
For spherically symmetric elastic matter there are 5 independent wave speeds~\cite{Karlovini:2002fc}. However, from these, only 3 can be obtained from a spherically symmetric stored energy function $\widehat{w}(\delta,\eta)$ (equivalently $\widehat{\rho}(\delta,\eta)$), namely the speed of \emph{longitudinal} waves in the \emph{radial} direction,
\begin{equation}
c^2_\mathrm{L}({{\delta}},\eta)= \frac{{{\delta}}\partial_{{\delta}} \widehat{p}_{\mathrm{rad}}}{\widehat{\rho}+\widehat{p}_{\mathrm{rad}}}\,,\\
\end{equation}
and the speeds of \emph{transverse} waves in the \emph{radial} and \emph{tangential} directions (the later oscillating in the \emph{radial} direction):
\begin{eqnarray}
	c^2_\mathrm{T}({{\delta}},\eta)&=& \frac{\widehat{p}_{\mathrm{tan}} - \widehat{p}_{\mathrm{rad}}}{\left(\widehat{\rho}+\widehat{p}_{\mathrm{tan}}\right)\left(1 - {{\delta}}^2/\eta^2\right)}\,, \\
\tilde{c}^2_\mathrm{T}({{\delta}},\eta)&=& \frac{\widehat{p}_{\mathrm{rad}} - \widehat{p}_{\mathrm{tan}}}{\left(\widehat{\rho}+\widehat{p}_{\mathrm{rad}}\right)\left(1 - \eta^2/{{\delta}}^2\right)}\,.
\end{eqnarray}
In general, the remaining 2 velocities, corresponding to \emph{longitudinal} waves in the \emph{tangential} direction ($\tilde{c}_\mathrm{L})$ and to \emph{transverse} waves in the \emph{tangential} direction oscillating in the \emph{tangential} direction ($\tilde{c}_{\mathrm{TT}}$), can only be obtained from a stored energy function without symmetries. 
However, as shown in~\cite{inprep}, these two velocities satisfy the relation
\begin{equation}\label{RelWS}
\tilde{c}^2_\mathrm{L}(\delta,\eta)-\tilde{c}^2_\mathrm{TT}(\delta,\eta)=\frac{\delta^2\partial^2_\delta\widehat{\rho}+3\delta\eta\partial^2_{\eta\delta}\widehat{\rho}+\frac{9}{4}\eta^2\partial^2_\eta \widehat{\rho}+\frac{3}{4}\eta\partial_\eta\widehat{\rho}}{\widehat{\rho}+\widehat{p}_\mathrm{tan}}.
\end{equation}
Therefore, spherically symmetric elastic materials are described by two functions, $\widehat{\rho}(\delta,\eta)$ and either $\tilde{c}^2_\mathrm{L}(\delta,\eta)$ or $\tilde{c}^2_{\mathrm{TT}}(\delta,\eta)$ (since these are related by the above constraint). Moreover, it is also shown in~\cite{inprep} that the simplest expression for $\tilde{c}^2_\mathrm{L}$ satisfying the \emph{isotropic state condition} at the center is given by
\begin{equation}
\tilde{c}^2_\mathrm{L}(\delta,\eta)=\frac{{{\delta}}\partial_{{\delta}} \widehat{p}_{\mathrm{tan}}+3\eta\partial_\eta \widehat{p}_{\mathrm{tan}}}{\widehat{\rho}+\widehat{p}_{\mathrm{tan}}}
\end{equation}
(the so-called \emph{natural choice}), so that the corresponding value of $\tilde{c}^2_\mathrm{TT}$ is
\begin{equation}
\tilde{c}^2_\mathrm{TT}(\delta,\eta)=\frac{\frac{3}{2}\eta\partial_\eta\widehat{p}_\mathrm{tan}}{\widehat{\rho}+\widehat{p}_\mathrm{tan}}.
\end{equation}
Overall, reality and causality require $0\leq {c}^2_\mathrm{L,T}\leq1$ and $0\leq \tilde{c}^2_\mathrm{L,T,TT}\leq1$.

\begin{figure}[t]
	\centering
	\label{fig:profiles}
 	\includegraphics[width=0.48\textwidth]{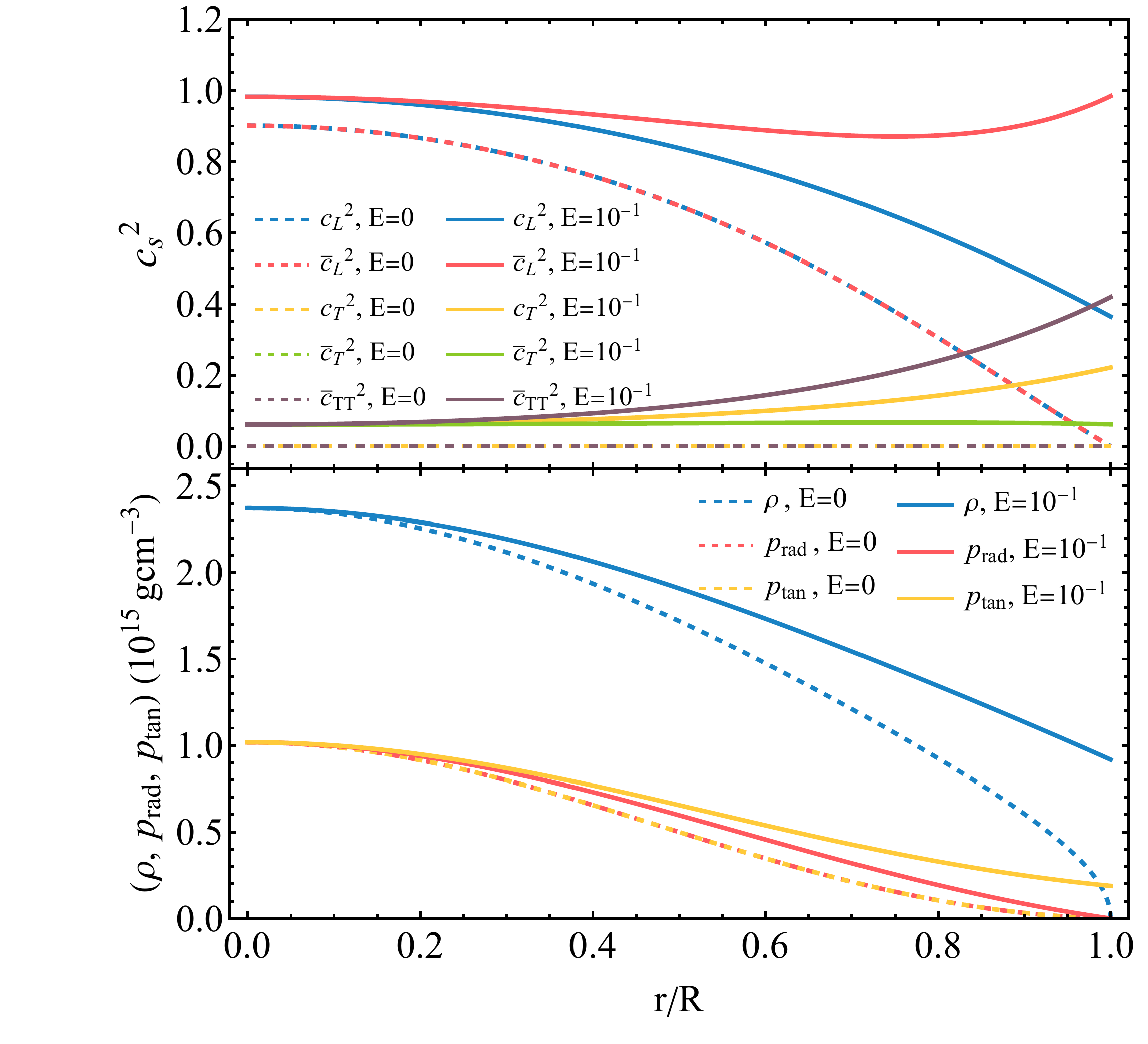} 
	\caption{
	Sound speeds (upper panel) and density and pressure profiles (bottom panel) for the quadratic elastic model with $n=1/2$, $K=6\times 10^4 M_\odot^{4}$. We compare the perfect fluid case ($E=0$, dashed lines) with an elastic configuration with $E=10^{-1}$ (solid lines). The latter configuration features a light ring ($M/R\approx0.35$) and the wave speeds are always subluminal. 
	}
\end{figure}

Figure~\ref{fig:MR} shows the $M-R$ diagram for some representative examples of the aforementioned elastic models. 
Larger values of $E$ and of $s-n$ increase the maximum mass and maximum compactness of star. In particular, a one-parameter deformation of the $n=1/2$ polytrope (left panel) supports configurations with a light ring ($R<3M$) and exceeding the Buchdahl's limit ($R<9M/4$).
However, a relevant question is whether the equilibrium configurations satisfy the various constraints imposed by the subluminality of wave propagation in the radial and tangential directions. Solid (dashed) parts of each curve in Fig.~\ref{fig:MR} correspond to subluminality (superluminality). Depending on the model, the transition can occur before or after the maximum mass, which corresponds to the heaviest stable configuration under radial perturbations, as discussed below.
In general, as the central density increases, some wave speeds (predominantly $\tilde{c}_\mathrm{L}$, i.e. the speed of longitudinal waves along the tangential direction) become superluminal within the star, as shown in Fig.~\ref{fig:profiles} for a configuration that is almost marginally causal.
%
These viability requirements set an upper bound on the maximum mass and compactness of physically realizable equilibrium solutions. Additionally, we find that all subluminal and stable solutions satisfy all the energy conditions.  In all models under consideration, the Buchdahl limit is exceeded only for radially unstable or superluminal configurations. On the other hand, we found physically realizable ultracompact configurations ($R<3M$) for both models~\eqref{QM} and \eqref{PolyStore2} with $n=1/2$, although the second case requires a fine tuning of the parameters and the compactness never exceeds $M/R=1/3$ by more than $1\%$~\cite{inprep}.

For the simple quadratic model~\eqref{QM} with $n=1/2$, we find stable and causal configurations with compactness as large as $M/R\approx 0.35$, which is reached for $E={\cal O}(10^{-1})$. Interestingly, in this case the shear modulus is approximately 
\begin{equation}
 \mu \approx 7\times 10^{26}\left(\frac{\rho_0}{10^{11}\,{\rm g/cm}^3}\right)^2\sqrt{\frac{K}{10^5 M_\odot^4}}\left(\frac{E}{0.1}\right)\,{\rm erg/cm}^3\,,
\end{equation}
which is in the typical range of values for lattice models describing the neutron-star crust~\cite{Chamel:2008ca}.

\noindent{{\bf{\em Radial stability.}}}
Linear radial perturbations of relativistic elastic balls have been treated in~\cite{Karlovini:2003xi} using a Lagrangian approach. In~\cite{inprep} we use our new Eulerian definition for spherically symmetric elastic bodies and linearise the Einstein equations around the static background,
also including perturbations of a (possibly non-flat) material metric. The main perturbation variables are $\zeta(t,r)\equiv r^2 e^{-\alpha}\xi$ and $\chi(t,r)=-e^{\beta+2\alpha}(p_\mathrm{rad})_\mathrm{L}$,
where $\xi$ is the usual radial displacement in the perturbed configuration and $(p_\mathrm{rad})_\mathrm{L}$ is the Lagrangian perturbation of the radial pressure.
Making the ansatz $\zeta(t,r)=e^{i  \omega t}\zeta(r)$, $\chi(t,r)=e^{i  \omega t}\chi(r)$
leads to an eigenvalue problem for the system of first-order ordinary differential equations
%
\begin{subequations}
	\begin{align}
	{{\delta}}\partial_{{\delta}}\widehat{p}_\mathrm{rad}\frac{d\zeta}{dr} &= -\frac{3}{r}\eta\partial_\eta\widehat{p}_\mathrm{rad}\zeta+e^{-(3\alpha+\beta)}r^2\chi \,,\\
	{{\delta}}\partial_{{\delta}}\widehat{p}_\mathrm{rad} \frac{d\chi}{dr}&=\frac{3}{r}\eta\partial_\eta\widehat{p}_\mathrm{rad}\chi-\left[Q_1+Q_2\omega^2\right]\zeta \,,
	\end{align}
\end{subequations}
where
\begin{subequations}
	\begin{align}
Q_1 &=\frac{e^{3\alpha+\beta}}{r^2}\Big[\frac{4}{r^2}(\delta\partial_\delta\widehat{q}-\widehat{q})^2+\delta\partial_\delta\widehat{p}_\mathrm{rad}\Big(\frac{2}{r^2}\widehat{q}-\frac{6}{r^2}\eta\partial_\eta\widehat{q} \nonumber\\
&-\frac{4}{r}\frac{dp_\mathrm{rad}}{dr}+\left(\frac{6}{r}\widehat{q}-\frac{dp_\mathrm{rad}}{dr}\right)\frac{d\alpha}{dr}-8\pi(\widehat{\rho}+\widehat{p}_\mathrm{rad})e^{2\beta}\widehat{p}_\mathrm{rad}\Big)\Big]\,, \\
	Q_2&=\frac{e^{\alpha+3\beta}}{r^2}\delta\partial_\delta\widehat{p}_\mathrm{rad}(\widehat{\rho}+\widehat{p}_\mathrm{rad})	\,,
	\end{align}
\end{subequations}
subject to the standard boundary conditions~\cite{Kokkotas:2000up} $\lim\limits_{r\rightarrow 0^+}\frac{\zeta}{r}=0$ and $\lim\limits_{r\rightarrow R}\chi=0$.

\begin{figure}[t]
	\centering
	\label{fig:stability}
 	\includegraphics[width=0.48\textwidth]{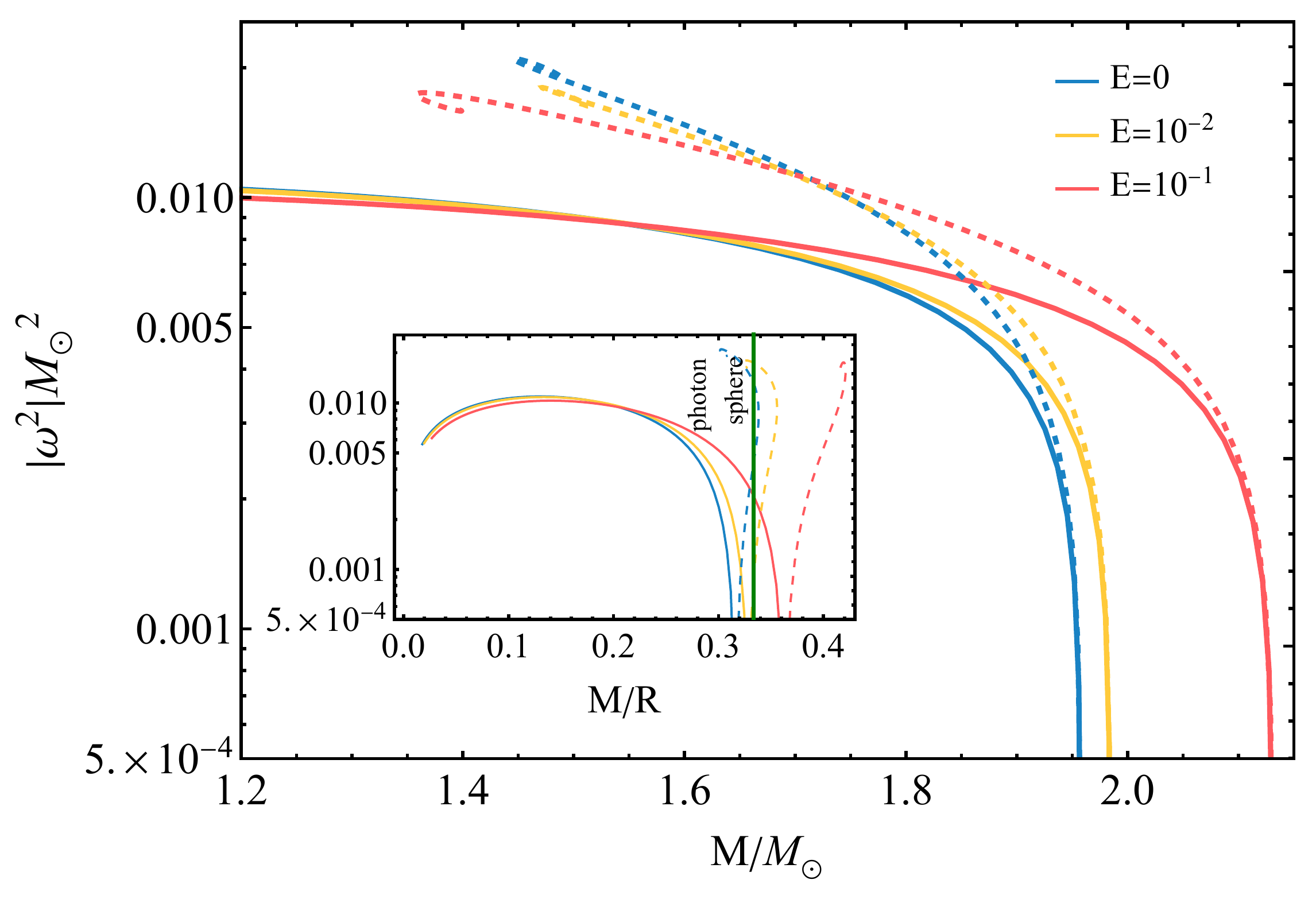} 
	\caption{Squared frequency eigenvalues for the radial stability analysis of elastic stars in the quadratic model~\eqref{QM} with $n=1/2$ and different values of $E$ as a function of the mass or of the compactness (see inset). Solid (dashed) lines correspond to stable (unstable) configurations with $\omega^2>0$ ($\omega^2<0$). In all cases the zero crossing corresponds to the maximum mass within numerical accuracy. 
	}
\end{figure}

As a representative example, in Fig.~\ref{fig:stability} we show the eigenvalues $\omega^2$ obtained numerically for model~\eqref{QM} as a function of the mass for $n=1/2$ and a range of values of $E$.  Within numerical accuracy, the zero crossing corresponds to the point of the maximum mass. We found the same result in all models under investigation~\cite{inprep}. Based on this numerical evidence, we conclude that elastic stars beyond the maximum-mass configuration are radially unstable, as in the perfect fluid case~\cite{Kokkotas:2000up}.

\noindent{{\bf{\em Discussion.}}}
%
Anisotropies are ubiquitous in physical systems and seem also a key ingredient to built solutions for ultracompact self-gravitating objects. However, introducing anisotropies in GR is often based on ad-hoc models which might also suffer from violation of the energy conditions or superluminal wave propagation. We presented a general framework to build pathology-free relativistic compact objects containing elastic materials.

All elastic relativistic configurations that we have constructed have ${p}_{\mathrm{tan}}\geq {p}_{\mathrm{rad}}$ within the star. Interestingly, this condition violates the mild-anisotropy assumption of the Buchdahl's theorem~\cite{Urbano:2018nrs}, and in fact elasticity allows for compact objects which can exceed the Buchdahl's limit, $M/R>4/9$. However, in all models we have explored this limit is reached only in the branch that is unstable against radial perturbations. In addition to radial stability, we advocate the importance of checking the (nontrivial) viability of the matter fields for anisotropic configurations, in particular \emph{all} causality conditions for wave propagation.
While these requirements limit the maximum mass and compactness of elastic stars, we showed that it is nevertheless possible to obtain physically realizable ultracompact configurations featuring a light ring.
Elastic stars are one of the few GR models of ultracompact objects that satisfy all the above viability requirements and come from a first-principles Lagrangian approach. The only other known example in the static case are quark stars~\cite{Mannarelli:2018pjb,Bora:2020cly,Urbano:2018nrs,Zhang:2021fla} that however require strange matter. 

Due to the photon sphere, nonradial perturbations of ultracompact elastic stars can feature gravitational-wave echoes~\cite{Cardoso:2016rao,Cardoso:2017cqb,Pani:2018flj}, as in the quark star case~\cite{Mannarelli:2018pjb,Bora:2020cly,Urbano:2018nrs,Zhang:2021fla}. 
Horizonless compact objects with an unstable light ring feature a second, stable, inner photon sphere, which might be prone to nonlinear instabilities~\cite{Cardoso:2014sna,Cunha:2017qtt,Ghosh:2021txu}. Investigating this problem in detail was so far hampered by the lack of physically realizable and first-principle solutions, but it can be done within our framework. Indeed, by solving the null geodesic equations we have confirmed that ultracompact elastic stars do feature a stable light ring.

Elasticity tends to support more massive and compact configurations. For the models at hand the maximum mass can increase up to $\approx 22\%$ relative to the perfect fluid case while the material remains physically realizable. Interestingly, this suggests that static neutron star models, that can reach $M\approx 2M_\odot$ in the perfect fluid case, could potentially be as massive as $M\approx 2.5M_\odot$ when elasticity is included, without violating any physical requirement. Such heavy neutron stars would be compatible with the exotic secondary object of the gravitational-wave event GW190814~\cite{Abbott:2020khf}.
Furthermore, more compact configurations would tend to have smaller tidal deformability. Thus, stiff EoS that are in tension with the relatively small tidal deformability measured by GW170817~\cite{GW170817} could evade those bounds when elasticity is included.

Further natural applications of our framework include considering: (i) other models of elastic materials, including non-flat reference material metrics; (ii) consistent multilayer solutions, e.g. made of a perfect fluid interior and an outer elastic crust~\cite{Chamel:2008ca,Suleiman:2021hre,Raposo:2020yjy}, or deformations of piecewise polytropes~\cite{Read:2008iy} that approximate tabulated, nuclear-physics based EoS; (iii) models that deform generic barotropic fluids beyond the polytropic EoS. We will report on these applications elsewhere~\cite{inprep}.
Finally, we focused here on spherical symmetry but our approach can be extended to less symmetric configurations. In particular, future work will also focus on rotating and (possibly tidally) deformed~\cite{Raposo:2020yjy} elastic solutions.

\noindent{{\bf{\em Acknowledgments.}}}
%
A.A. and J.N. were partially supported by FCT/Portugal through CAMGSD, IST-ID, projects UIDB/04459/2020 and UIDP/04459/2020.
P.P. acknowledges financial support provided under the European Union's H2020 ERC, Starting 
Grant agreement no.~DarkGRA--757480, and under the MIUR PRIN and FARE programmes (GW-NEXT, CUP:~B84I20000100001), and support from the Amaldi Research Center funded by the MIUR program `Dipartimento di Eccellenza" (CUP:~B81I18001170001). 
G.R. was supported by FCT/Portugal through  the grant PTDC/FIS-OUT/28407/2017 and the Center for Research and Development in Mathematics and Applications (CIDMA), projects UIDB/04106/2020 and UIDP/04106/2020. 

\bibliographystyle{utphys}
\bibliography{biblio}

\end{document}